\documentclass[aps,prl,twocolumn,superscriptaddress,floatfix,amsmath,showpacs]{revtex4}

\usepackage{graphicx}

\begin{document}

\preprint{APS/123-QED}

\newcommand{\Hb}{\bf$\bigotimes$}

\title{Direct current control of three magnon scattering processes in spin-valve nanocontacts}

\author{H.~Schultheiss}
\affiliation{Fachbereich Physik and Forschungszentrum OPTIMAS, Technische Universit\"{a}t Kaiserslautern, 67663 Kaiserslautern, Germany}

\author{F.~Ciubotaru}
\affiliation{Fachbereich Physik and Forschungszentrum OPTIMAS, Technische Universit\"{a}t Kaiserslautern, 67663 Kaiserslautern, Germany}

\author{A.~Laraoui}
\affiliation{Fachbereich Physik and Forschungszentrum OPTIMAS, Technische Universit\"{a}t Kaiserslautern, 67663 Kaiserslautern, Germany}

\author{S.J.~Hermsdoerfer}
\affiliation{Fachbereich Physik and Forschungszentrum OPTIMAS, Technische Universit\"{a}t Kaiserslautern, 67663 Kaiserslautern, Germany}

\author{B.~Obry}
\affiliation{Fachbereich Physik and Forschungszentrum OPTIMAS, Technische Universit\"{a}t Kaiserslautern, 67663 Kaiserslautern, Germany}

\author{A.A.~Serga}
\affiliation{Fachbereich Physik and Forschungszentrum OPTIMAS, Technische Universit\"{a}t Kaiserslautern, 67663 Kaiserslautern, Germany}

\author{X.~Janssens}
\affiliation{IMEC, Kapeldreef 75, 3001 Leuven, Belgium}

\author{M.~van Kampen}
\affiliation{IMEC, Kapeldreef 75, 3001 Leuven, Belgium}

\author{L.~Lagae}
\affiliation{IMEC, Kapeldreef 75, 3001 Leuven, Belgium}

\author{A.N.~Slavin}
\affiliation{Department of Physics, Oakland University, Rochester, MI-48309, USA}

\author{B.~Leven}
\affiliation{Fachbereich Physik and Forschungszentrum OPTIMAS, Technische Universit\"{a}t Kaiserslautern, 67663 Kaiserslautern, Germany}

\author{B.~Hillebrands}
\affiliation{Fachbereich Physik and Forschungszentrum OPTIMAS, Technische Universit\"{a}t Kaiserslautern, 67663 Kaiserslautern, Germany}

\begin{abstract}
We have investigated the generation of spin waves in the free
layer of an extended spin-valve structure with a nano-scaled point
contact driven by both microwave and direct electric current
using Brillouin light scattering microscopy. Simultaneously with the directly excited spin waves, strong nonlinear effects are observed, namely the generation of eigenmodes with integer multiple frequencies (2\,\emph{f}, 3\,\emph{f}, 4\,\emph{f}) and modes with non-integer factors (0.5\,\emph{f}, 1.5\,\emph{f}) with respect to the excitation frequency \emph{f}. The origin of these nonlinear modes is traced back to three magnon scattering processes. The direct current influence on the generation of the fundamental mode
at frequency \emph{f} can be related to the spin-transfer torque,
while the efficiency of three-magnon-scattering processes is
controlled by the Oersted field as an additional effect of the
direct current.

\end{abstract}


\maketitle

Magnetic excitations generated by a high-density electric current
in a ferromagnetic nanostructure have attracted growing attention
due to the importance of understanding the physical
mechanisms responsible for the excitation process. The discovery
of the spin transfer torque (STT) effect in 1996 by Slonczewski
\cite{Slonczewski} and Berger \cite{Berger}, i.e. the excitation
of the precession of the magnetization in ferromagnetic thin films
and nanostructures by a spin polarized direct electric current,
has offered a new scheme for spin wave excitation in magnetic
nanostructures. In a point contact structure, this torque may excite
spin waves whose frequencies are tunable via
both current and applied magnetic field \cite{Stiles, Rippard}.
The electrical contacts made with diameters less than 100\,nm to a
continuous spin-valve multilayer stack permit to achieve huge
current densities with relatively small applied currents. If the
current density is large enough, the STT compensates the natural
dissipation processes. This may lead to self-sustained dynamics under and
near the point contact such as the formation of nonlinear
evanescent bullet modes in the case of in-plane magnetization
\cite{Slavin, Tiberkevich}. The impact on spin waves by a microwave current flowing through a point contact was demonstrated \cite{Russek, Katine}, with possible
applications in, e.g., radio-frequency devices for wireless
communication.

\begin{figure}[t]
\begin{center}
\includegraphics[width=8.5 cm, clip]{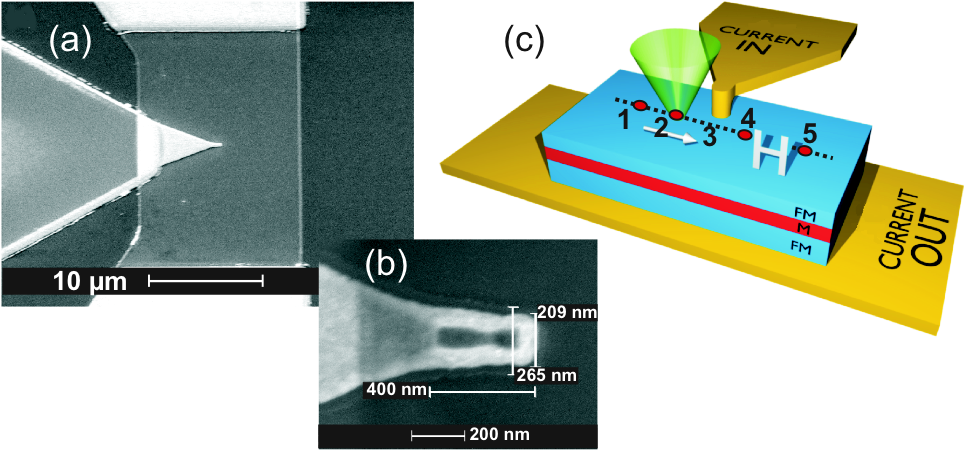}
\end{center}
\vspace*{0 cm}\caption{SEM images of the sample (a) and of the top
electrode (b). (c) Schematic representation of the sample with
focused light near the point contact in the presence of an
external in-plane bias magnetic field \emph{H}.} \label{Fig1}
\end{figure}

In this letter we report on spin-wave excitations in spin-valve nanocontacts by means of Brillouin light scattering (BLS) microscopy. This technique uses a focused laser spot to probe the spin waves with a spatial resolution of 250\,nm and a frequency resolution of up to 50\,MHz. A detailed description of the experimental setup can be found in \cite{Demidov, Perzlmaier}. As a result of the very high current densities which can be achieved in a nanocontact, the spin wave system can be driven to amplitudes far beyond the linear regime. Hence, nonlinear processes such as the splitting and the confluence of spin waves can occur. The influence of an additionally applied direct current on these nonlinear processes, either due to STT or the Oersted field, is investigated.

The investigated structures consist of an extended spin-valve multilayer sequence (IrMn(6\,nm)/\-Co$_{90}$Fe$_{10}$(5\,nm)/\-Cu(3.5\,nm)/\-Ni$_{80}$Fe$_{20}$(7\,nm)/\-Pt(3\,nm)) with lateral dimensions of 15$\times$45$\,\mu$m$^{2}$. The electric current is injected through a nano-scaled metallic point contact on top of the spin valve. The Ni$_{80}$Fe$_{20}$ layer acts as the magnetic free layer of the spin-valve and the Co$_{90}$Fe$_{10}$ layer, which is pinned by an antiferromagnetic layer (IrMn), serves as a reference layer for the giant magneto-resistance (GMR) measurements. The spin-valve is sputter-grown in an ultrahigh vacuum system with a base pressure of 3$\cdot$10$^{-8}$\,Torr. After patterning by conventional lift-off lithography, the stack is passivated by 50\,nm of SiO$_{2}$ for electric insulation. The point contact is then created by etching a hole with a diameter smaller than 120\,nm through the insulating SiO$_{2}$ layer. The leads are defined in a single e-beam lithography step. A feature of the investigated structure is the triangularly shaped top electrode. This configuration provides optical access to a large part of the magnetic area and enables the investigation of the magnetization dynamics close to the nanocontact. Fig.~\ref{Fig1} shows a scheme of the sample geometry and scanning electron microscopy (SEM) images of the point contact. Previous studies \cite{Kampen} of the same device have shown a very small coercive field  ~280\,A/m (3.5\,Oe) of the free magnetic layer. A unidirectional pinning field of 1.59\,kA/m (20 \,Oe) is most probably caused by RKKY or N\'eel coupling, but is negligible for the results presented here. The total resistance of the structure (19.4\,$\Omega$) and the resistance difference between the parallel and antiparallel alignment of the magnetic layers (48\,m$\Omega$) lead to a GMR ratio of 2.4\,$\%$ at room temperature.

\begin{figure}[t]
\begin{center}
\includegraphics[width=8.2 cm,clip]{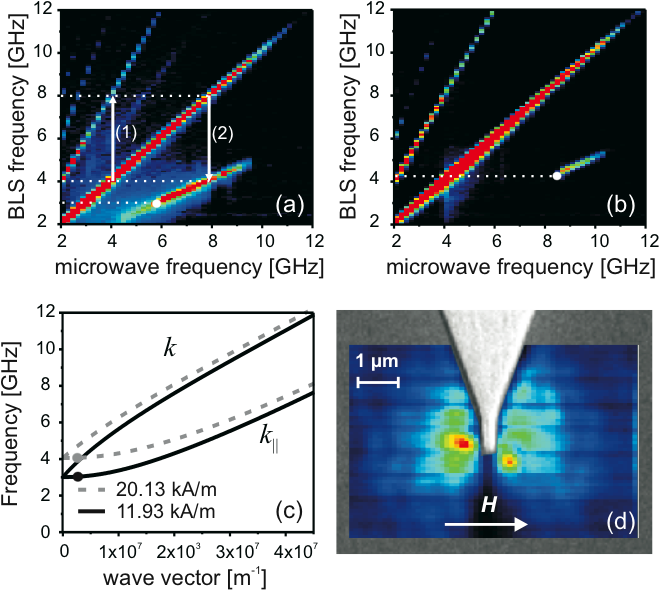}
\end{center}
\vspace*{0 cm}\caption{Color-coded spin-wave amplitude measured
with BLS-microscopy as a function of the RF excitation frequency
(x-axis) and the position in the BLS spectrum (y-axis) for an
in-plane magnetic field of (a) 11.93\,kA/m (150\,Oe) and (b)
20.13\,kA/m (250\,Oe). Black indicates small precession angles
and red indicates large precession angles. (c) Spin-wave
dispersion band for the magnetic fields corresponding to panels
(a) and (b). The top and bottom branches correspond to wave
vectors oriented perpendicular and parallel to the magnetic field.
(d) BLS microscopy image of the two-dimensional distribution of the spin waves radiated from
the contact at a pumping frequency of 3.1\,GHz.} \label{Fig2}
\end{figure}

We excite spin waves in the structure by applying a microwave current with various frequencies (2 to 12\,GHz) and powers (0.1 to 20\,mW). Using a bias-tee, an additional direct current can be injected. Typical current densities used in our experiments range from 10$^{11}$ to 10$^{12}$\,A/m$^2$. The sample is saturated by an external magnetic field \emph{H} applied in the plane of the magnetic layer, parallel to the long axis of the spin valve stack (Fig.~\ref{Fig1}c). In the first step we characterize the dynamic response of the device as a function of the applied frequency in order to determine the resonance frequencies of the Ni$_{80}$Fe$_{20}$ layer close to the point contact. The results are shown in Fig.~\ref{Fig2}a and \ref{Fig2}b for externally applied fields of 11.93\,kA/m (150\,Oe) and 20.13\,kA/m (250\,Oe), respectively. These intensity graphs display the measured BLS intensity in a color code as a function of the microwave excitation frequency (x-axis) and the frequency position in the BLS spectrum (y-axis). The diagonal in the figure re\-pre\-sents the efficiency of the excitation of spin waves with frequencies that match the applied microwave frequency \emph{f}. Simultaneously with the directly excited spin waves, strong nonlinear effects appear, namely the generation of modes with integer multiple frequencies (2\,\emph{f}, 3\,\emph{f}, 4\,\emph{f}) and modes with non-integer factors (0.5\,\emph{f}, 1.5\,\emph{f}) with respect to the excitation frequency.

The excitation of these nonlinear modes can be understood within the framework of three-magnon scattering processes \cite{Melkov} in which the total energy and momentum are conserved. The white arrows in Fig.~\ref{Fig2}a show directly the conservation of the
energy in those processes. In the case of confluence, two magnons with frequency \emph{f} combine into a single magnon with doubled frequency 2\,\emph{f} (process 1 in Fig.~\ref{Fig2}a). In the other case, one magnon with frequency \emph{f} splits in two
magnons, each one having half the frequency \emph{f}/2 (process 2). From Figs.~\ref{Fig2}a and \ref{Fig2}b it is evident that the splitting process does not appear below a certain threshold frequency (white dots in Figs.~\ref{Fig2}a,b), which depends on the applied field. With an increasing magnetic field this threshold frequency increases as well from approximately 3\,GHz for $H$=11.93\,kA/m (150\,Oe) to $~$4.3\,GHz for $H$=20.13\,kA/m (250\,Oe).

This threshold frequency can be derived from the analytical dispersion relations of spin-wave eigenmodes in a 7\,nm thick Ni$_{80}$Fe$_{20}$ layer. We calculated the dispersion relations following the equations in \cite{Kalinikos1986} and plotted them in Fig.~\ref{Fig2}c for the corresponding values of the applied magnetic fields (11.93\,kA/m and 20.13\,kA/m). We used standard material parameters for the saturation magnetization ($M_{\textrm{s}}$ =800\,kA/m) and the exchange constant ($A$ =1.3$\cdot$10$^{-11}$\,J/m). Note that the dispersion of spin waves in a thin film is strongly anisotropic, depending on the propagation direction of spin waves with respect to the static magnetization vector. For each magnetic field, the upper (lower) curve corresponds to spin waves propagating perpendicular (parallel) to the magnetization vector, respectively. These two extremal dispersion curves define the frequency band in which spin-wave eigenmodes are allowed. Therefore, taking energy conservation into account, it is expected that a splitting process can only occur if the excitation frequency is at least twice the frequency at the bottom of the spin-wave band. The observed threshold frequencies for the \emph{f}/2 modes match the frequency at the bottom of the spin-wave band (white dots in the Fig.~\ref{Fig2}c). This supports the interpretation of three magnon scattering as the responsible mechanism for the generation of the \emph{f}/2 modes.

A typical radiation pattern of spin waves in the Ni$_{80}$Fe$_{20}$ is shown in Fig.~\ref{Fig2}d. In this case a frequency of 3.1\,GHz was applied at a magnetic field of $H = 11.93$\,kA/m. It is clearly visible that no spin waves with wave vectors perpendicular to the applied field are created. Thus the radiation pattern inherits the same anisotropy as the dispersion relations.

\begin{figure}[t]
\begin{center}\includegraphics[width=8.5 cm,clip]{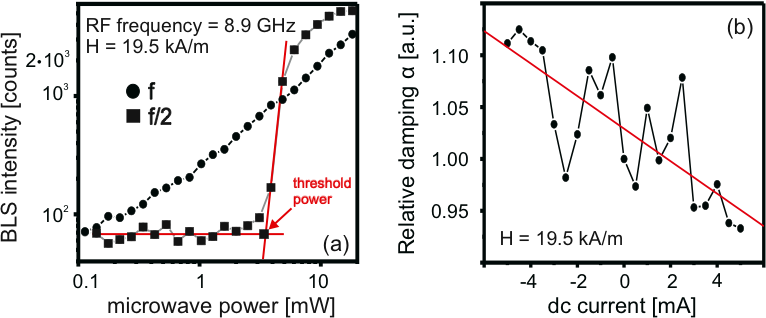}
\end{center}
\vspace*{0 cm}\caption{(a) BLS intensity of the $\emph{f}$ and
\emph{f}/2 modes as a function of microwave excitation power (x-axis).
The indicated point represents the threshold power for the \emph{f}/2
mode. (b) Relative damping of the directly excited spin waves at
8.9\,GHz as a function of the direct current. Line represents the
best linear fit of the data.}
\label{Fig3}
\end{figure}

Since the splitting of magnons is a nonlinear process, the amplitude of the excited spin waves (\emph{f} and \emph{f}/2) was studied as a function of the applied microwave power (Fig.~\ref{Fig3}a). The excitation frequency is fixed at 8.9\,GHz, just above the threshold frequency which was found for the splitting process in Fig.~\ref{Fig2}b. As expected, the amplitude of the directly excited spin waves increases linearly with the applied microwave power. The \emph{f}/2 mode, however, shows a very different dependency. Starting from very low microwave powers it cannot be detected over a wide range. Upon crossing a certain threshold power, the amplitude increases drastically, and even overcomes the amplitude of the directly excited spin waves. This threshold behavior is characteristic for three magnon scattering processes.

The absolute value of the threshold power for the splitting of magnons is a function of many parameters \cite{Melkov}. Most important for the studies presented here is the magnitude of the internal magnetic field and the damping parameter of the spin waves. Since it is well known that spin transfer torque can compensate the natural damping of spin waves, it is of particular interest to investigate the influence of a spin polarized direct current on the \emph{f} and \emph{f}/2 modes. In the experimental geometry a positive polarity of the direct current corresponds to a flow of electrons from the fixed layer to the free layer. For a fixed applied microwave power, the amplitude of the \emph{f} mode was measured as a function of the direct current. Demokritov \emph{et al.} showed in \cite{Demokritov} that $\alpha$ is proportional to the inverse square root of the BLS intensity. Following their analysis
with the experimental results reported here a decrease of the Gilbert damping parameter was observed ($\alpha$) with increasing direct current (Fig.~\ref{Fig3}b), consistent with the STT effect. Note that in Fig.~\ref{Fig3}b the values of $\alpha$ are normalized to the case of zero applied current.

\begin{figure}[t]
\begin{center}
\includegraphics[width=8.5 cm,clip]{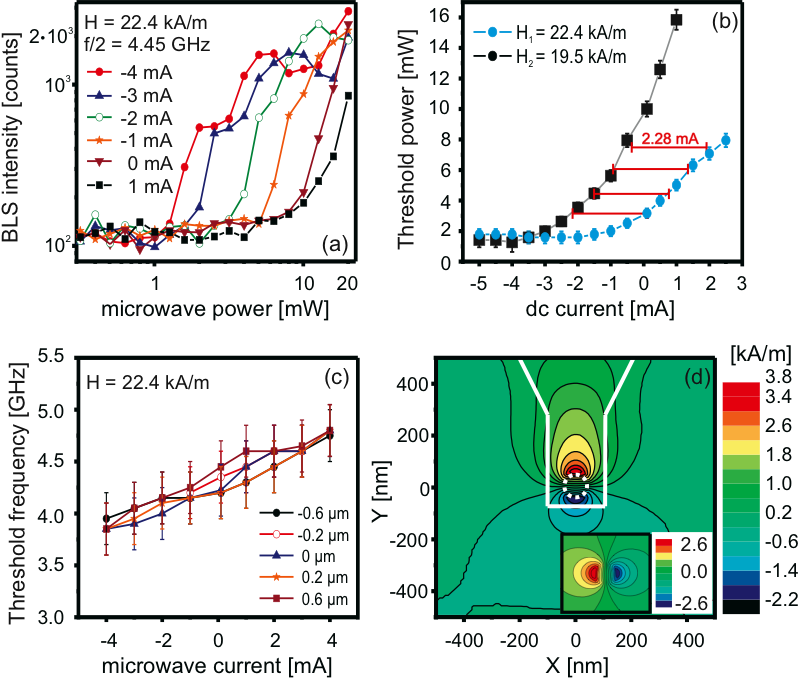}
\end{center}
\vspace*{0 cm}\caption{(a) Generation efficiency of the \emph{f}/2
mode (=4.45\,GHz) as a function of the applied microwave power for
different direct currents ($H$=22.4\,kA/m). (b)
Extracted threshold power for the \emph{f}/2 mode over the direct
current for two values of the applied field. (c) Influence of the
scan position on the threshold frequency over the direct
current for 20\,mW. (d) Spatial distribution of the Oersted field components
parallel and perpendicular (inset) to the applied field. The white
continuous line represents the border of the top electrode while
the dotted white curve delimits the edge of the contact. The inset
figure has the lateral size of 300\,nm $\times$ 300\,nm.}
\label{Fig4}
\end{figure}

Furthermore, a strong influence of the direct current on the power dependency of the splitting processes was observed and is shown in Fig.~\ref{Fig4}a. The threshold power for the excitation of the \emph{f}/2 mode is growing by orders of magnitude with rising direct currents. Fig.~\ref{Fig4}b shows the extracted threshold powers as a function of the direct current for two values of the applied magnetic field (19.5\,kA/m (245\,Oe) and 22.4 kA/m (281\,Oe)). A nonlinear relation between the threshold power and the direct current is apparent. From theory it is known that the STT causes a linear relation between $\alpha$ and the direct current, leading to a linear response of the threshold power as a function of the current strength \cite{Slonczewski, Buhrman}. Obviously, this linear behavior is not observed in the experimental data presented here. Moreover, the increase of the threshold power with increasing direct current suggests that the internal damping should get larger as well. These results are in contradiction with a STT-based mechanism.

So far the influence of the Oersted field created by the direct current was not addressed. An additional magnetic field created by the current will change the dispersion relations of the spin waves. In particular, it will modify the frequency at the bottom of the spin-wave band. For this reason, the connection between the threshold frequency of the \emph{f}/2 mode and the magnitude of the direct current as well as the distance to the point contact was studied. The results, plotted in Fig.~\ref{Fig4}c, show that the direct current induces a proportional shift of the threshold frequency of the \emph{f}/2 mode, very similar to the results produced by changing the applied magnetic field (Fig.~\ref{Fig2}a,b). We determined the contribution of the Oersted field by adding an additional magnetic field to the dispersion relation. This field is chosen so that the bottom frequency of the spin wave band matches the threshold frequency observed experimentally in Fig.~\ref{Fig4}c. This analysis yields an additional magnetic field of 1.27\,(kA/m)/mA (16\,Oe/mA) created by the direct current.

Using a micromagnetic code \cite {Scheinfein} the current distribution and the corresponding Oersted field were simulated in the free layer with and without the contribution of the top electrode, respectively. It was found that the lead induces an additional magnetic field in the x direction caused by the Oersted field of the flowing direct current, as can be seen in Fig.~\ref{Fig4}d. This additional magnetic field is proportional to the applied direct current yielding a contribution of 1.2 (kA/m)/mA ($\sim$ 15\,Oe/mA). This value is in very good agreement with the magnetic field that we used to match the spin-wave dispersion to the obtained experimental results in Fig.~\ref{Fig4}c. Hence, the inhomogeneous magnetic field within the nanocontact area is responsible for the behavior of the frequency and power thresholds of the \emph{f}/2 when the direct current is changed. Note that in Fig.~\ref{Fig4}b the threshold power for the generation of the \emph{f}/2 mode for the magnetic fields $H_{1}$ and $H_{2}$ are shifted by a constant value of 2.3\,mA with respect to each other. This current difference corresponds to a magnetic field of 2.9\,kA/m (36\,Oe) using the calibration factor 1.27\,(kA/m)/mA (16\,Oe/mA) for the Oersted field created by the direct current. This value matches exactly the difference $\Delta H=H_{1}-H_{2}$ in both applied magnetic fields.

Clear evidence for the generation of the \emph{f}/2 mode \emph{inside} the contact arises from the measurements made for different scan positions as indicated in Fig.~\ref{Fig1}c. The results for different positions shown in Fig.~\ref{Fig4}c prove that the threshold frequency does not depend on the distance from the contact. Moreover, the threshold power for the \emph{f}/2 mode does not depend on the scan position. These results suggest that the Oersted field created outside the contact, which is strongly position dependent, has no contribution to the splitting process of the magnons. The magnons with half the excitation frequency are generated inside the contact area. Due to the very low group velocity of spin waves close to the bottom of the spin-wave band the \emph{f}/2 mode remains localized close to the nanocontact.

In conclusion, the magnetization dynamics in spin-valve nanocontacts has been investigated under the influence of an microwave and direct current by means of Brillouin light scattering microscopy. The huge current densities that can be achieved in a nanocontact allow the excitation of spin waves in the nonlinear regime. In the presence of a direct current the efficiency of the excitation of spin waves can be enhanced or reduced due to the spin transfer torque effect. The experimental results reported here show that magnon splitting and confluence processes take place within the nanocontact. The threshold properties of spin waves with half the excitation frequency are controlled by the Oersted field created by the current within the point contact and due to the Oersted field contribution of the asymmetric top electrode.

This work was supported by the European Commission within the EU-MRTN SPINSWITCH (MRTN-CT-2006-035327) and by the Deutsche Forschungsgemeinschaft within the SPP1133.

\end{document}